\pdfoutput=1
\documentclass[aps,prd,reprint,superscriptaddress,longbibliography,nofootinbib]{revtex4-2}

\usepackage{graphicx}
\usepackage{amsmath}
\usepackage{amssymb}
\usepackage{amsfonts}
\usepackage{dcolumn}
\usepackage{dsfont}
\usepackage{latexsym}
\usepackage{rotating}
\usepackage{color}
\usepackage{latexsym}
\usepackage{bbm}
\usepackage{subfigure}
\usepackage{float}
\usepackage{epsfig}
\usepackage{psfrag}
\usepackage{natbib, hyperref}
\usepackage{verbatim}
\usepackage{bm}
\usepackage{amsthm}
\usepackage{eucal}
\usepackage{mathrsfs}
\usepackage{url}
\usepackage{braket}
\usepackage{ulem}
\usepackage{calligra}
\usepackage[T1]{fontenc}
\usepackage[utf8]{inputenc}
\usepackage{newunicodechar}
\usepackage{marvosym}
\usepackage[absolute]{textpos}
\usepackage[cal=cm]{mathalfa}
\usepackage{soul}
\usepackage{appendix}

\usepackage{color}

\usepackage{hyperref}
\hypersetup{
colorlinks=true,final=true,
        linkcolor=blue,
        citecolor=blue,
        filecolor=blue,
        urlcolor=blue,
}

\begin{document}
\setlength{\abovedisplayskip}{3pt}
\setlength{\belowdisplayskip}{3pt}

\title{Superconducting pairing symmetries in charge-ordered kagom\'e metals}

\author{Pujita Das}
\affiliation{Department of Physics, Indian Institute of Technology Roorkee, Roorkee 247667, India}
\author{Parth Bahuguna}
\affiliation{Department of Physics, Indian Institute of Technology Roorkee, Roorkee 247667, India}
\author{Tulika Maitra}
\affiliation{Department of Physics, Indian Institute of Technology Roorkee, Roorkee 247667, India}
\author{Narayan Mohanta}
\email[]{narayan.mohanta@ph.iitr.ac.in}
\affiliation{Department of Physics, Indian Institute of Technology Roorkee, Roorkee 247667, India}

\begin{abstract}
We investigate the superconducting state in a kagom\'e lattice, with intertwined charge order and time-reversal symmetry-breaking loop current, using self-consistent Bogoliubov-de Gennes formalism to find the emergent pairing symmetries. Using local and nearest-neighbor attractive interactions, treated within Hartree-Fock mean-field approximation, we obtain all possible pairing symmetries in position space. Our findings indicate that the uniform $s$-wave symmetry, arising in the absence of the charge order and the loop current, modifies to a pair density wave of $s$-wave symmetry of 2$\times$2 lattice periodicity in the presence of the charge order, and a chiral pair density wave of $d_{x^2-y^2}\!+\!id_{xy}$-wave symmetry of the same 2$\times$2 periodicity in the presence of the charge order and loop current order, in both onsite and nearest-neighbor channels. In the absence of inversion symmetry, such as in the thin-film geometry, Rashba spin-orbit coupling appears, inducing an additional nearest-neighbor triplet $p_x\pm ip_y$-wave pairing. The results are relevant to superconductivity found in $A$V$_{3}$Sb$_{5}$ ($A$ = K, Rb, Cs), coexisting with a charge order that breaks time-reversal symmetry. We discuss fingerprints of these different pairing symmetries in scanning tunneling microscopy experiments.
\end{abstract}

\maketitle
\section{Introduction}
\vspace{-2mm}
Electronic instabilities in the metallic compounds $A$V$_{3}$Sb$_{5}$ ($A$ \!=\! K, Rb, Cs), including superconductivity below a critical temperature of 2.5 K and successive evolution of various density wave orders below a critical temperature of 94 K, have spurred their investigations in recent years~\cite{Ortiz_PRMat2019,Ortiz_PRL2020,Chen_Nat2021,Zhao_Nat2021,Jiang_NMat2021,Wang_PRR2021,Jiang_NaScRev2022,Li_NPhys2022,Shi_NCommun2022,Zhu_PRB2022,Lou_PRL2022,Guguchia_NCommun2023,Kang_NMat2023,Ge_PRX2024,Chen_PRMat2024}. The density-wave ordered states, which have been observed in experiments or theoretically proposed, are charge density wave, bond density wave, pair density wave and spin-density wave; many of these orders are generated spontaneously due to coupled degrees of freedom~\cite{Tan_PRL2021,Li_PRX2021,Kato_CommunMat2022,Wu_PRB2022,Song_ScChPhMeAstron2022,Guo_NPhys2024,Deng_PRB2025}. Muon spin rotation microscopy studies revealed broken time-reversal symmetry in the charge-ordered states, although without any long-range magnetic order, indicating the presence of a loop current flowing in the underlying lattice~\cite{Khasanov_PRR2022,Mielke_Nat2022,Shan_PRRes2022,Liege_PRB2024}. The two-dimensional kagom\'e lattice, formed by the V atoms, supports a flat dispersion-less band and Dirac points in the band structure and Van Hove singularities in the density of states. These intriguing electronic properties produce strong electronic-correlation effects and a wide variety of Berry phase-driven phenomena such as anomalous Hall effect, Nernst effect and magneto-Seebeck effect~\cite{Yang_sciadv2020, Gan_PRB2021, Yu_PRB2021, Zhou_PRB2022}. Because of the complex properties of the electronic bands and competing quantum phases, this family of compounds is a natural testbed for investigating unforeseen physical properties.

The superconducting state in $A$V$_{3}$Sb$_{5}$ has revealed many puzzling behaviors, quite similar to the high-temperature cuprate superconductors, such as the presence of a pseudogap-like precursor phase above the superconducting transition, coexistence with a competing charge order, electronic nematicity and a chiral flux phase order~\cite{Jin_PRL2021, Yu_NCommun2021, Xiang_NCommun2021, Yu_PRB2021, Feng_SciBull2021, Feng_PRB2021, Nie_Nat2022, Wulferding_PRRes2022, Neupert_NPhys2022}. The mechanism of electron pairing is not understood yet. There are some proposals for unconventional origins including electron-electron repulsive interaction, fluctuation in charge or spin degree of freedom and fluctuation in bond order~\cite{Romer_PRB2022,Chen_PRL2022,Yang_PRB2023,Tian_PRB2025, Chin_ChPhysLett2025}. However, growing evidences such as a sudden decrease in the splitting of Knight shift at the critical temperature $T_c$, existence of the Hebel-Slichter coherence peak near the $T_c$ and the existence of an orbital-selective electron-phonon interaction suggest conventional phonon-mediated pairing mechanism~\cite{Mu_ChPhLett2021, Xie_PRB2022, Roppongi_NCommun2023, Ritz_PRB2023}. A chiral pair density wave of Cooper pairs, induced by the loop-current in the coexisting charge-ordered state, is shown theoretically to reproduce some features of the pseudogap, V-shaped density of states and to predict an enhanced critical field for superconducting transition beyond the Pauli limit~\cite{Mohanta_PRB2023}. The formation of a chiral pair density wave state with broken time-reversal symmetry is consistent with the findings of scanning tunneling microscopy and muon spin rotation microscopy experiments~\cite{Chen_Nat2021}. The pressure-induced re-entrant superconducting phase when the charge order is suppressed indicates the realization of a non-trivial quantum state at low temperatures~\cite{Yu_NCommun2021}. Moreover, it is not established what kind of pair symmetries would emerge in the superconducting state in the presence of the charge order with broken time-reversal symmetry.

In this work, we consider a kagom\'e lattice with onsite and nearest-neighbor attractive interactions, a charge order modulation of $2a\!\times \!2a$ periodicity and a loop current order to describe the superconducting state in the compounds $A$V$_{3}$Sb$_{5}$. We use the Hartree-Fock mean-field approximation to treat the attractive interaction terms, and self-consistently compute pairing amplitudes in the onsite and nearest-neighbor channels. Guided by the irreducible representations (irreps), we identify all symmetry-allowed pairing symmetries in the absence and presence of the charge and loop current orders. We obtain pair density waves of $s$-wave Cooper pairs in the presence of only charge order modulation and chiral pair density wave of $d_{x^2-y^2}\!+\!id_{xy}$-wave Cooper pairs when both the charge order and loop current order are present in both onsite and nearest-neighbor channels. Both these pair density wave orders have the same $2\!\times \!2$ lattice periodicity, as in the charge order. We also consider the Rashba spin-orbit coupling which arises in the thin-film or interface geometry, and obtain a dominant triplet $p_x\pm ip_y$-wave pairing symmetry in the nearest-neighbor channel. We discuss relevance of our calculated results to available experimental findings and make some testable predictions for future detection of the found pairing symmetries.

The rest of the paper is organized as follows. In Sec.~\ref{secII}, we introduce our theoretical model and present the formalism that we used in our calculations with all related parameters. In Sec.~\ref{secIII}, we discuss possible pairing symmetries in the absence and presence of the time-reversal symmetry-breaking charge order. In Sec.~\ref{secIV}, we present our numerical results and discuss the influence of charge order, loop current and Rashba spin-orbit coupling. In Sec.~\ref{secV}, we show the local density of states in different realized pairing states and discuss their signatures in the scanning tunneling microscopy experiments. Finally, in Sec.~\ref{secVI}, we summarize our results and discuss future directions for the detection of the discussed pairing symmetries. In Appendix~\ref{App_A}, we present schematic real-space organization of superconducting order parameters belonging to the irreps in the point group $C_{6v}^{'''}$, describing the superconducting state in the presence of the charge and loop current orders.

\section{Theoretical Set Up}\label{secII}
\begin{figure}[t]
\includegraphics[width=83mm]{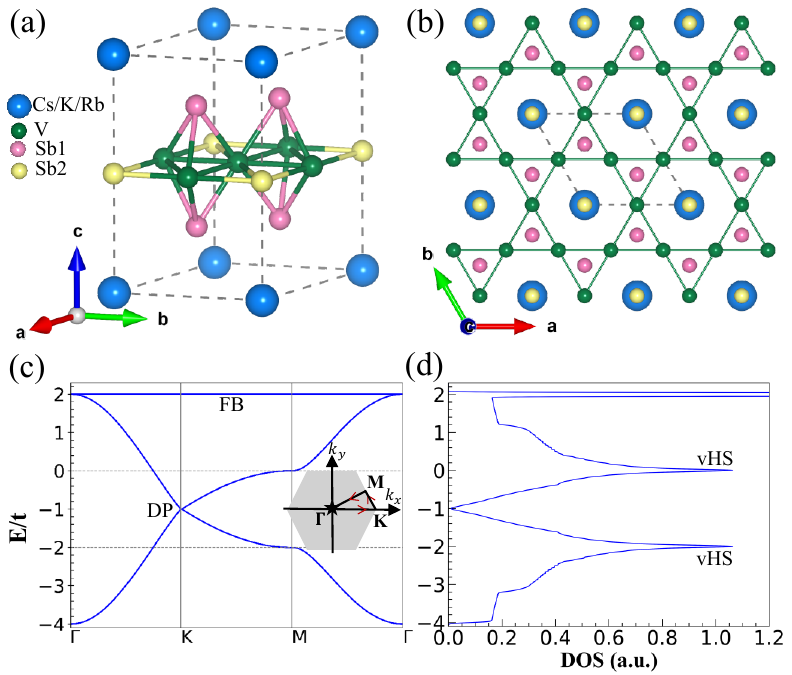}
\caption{(a) A unit cell of the compound $A$V$_{3}$Sb$_{5}$ ($A$ = Cs, K, Rb). (b) Top-down view of the crystal structure along the $c$ axis. The alkali atoms at the $A$ site are intercalated between V and Sb layers. The V atoms form a structurally perfect 2D kagom\'e lattice which is a triangular Bravais lattice with three inequivalent sublattices in a unit cell. There are two inequivalent Sb atoms: the in-plane Sb (yellow) atoms are located exactly at the center of the kagom\'e hexagons and the out-of-plane Sb (pink) atoms are located directly above and below the center of the equilateral triangles of the kagom\'e lattice. (c) Electronic band structure in a kagom\'e lattice, showing the Dirac points (DPs) and the flat band (FB). (d) The corresponding density of electronic states, showing the Van Hove singularities (VHSs) and the divergence near the flat band.}
%\vspace{-0mm}
\label{Fig1}
\end{figure}

\vspace{-2mm}
The compounds under consideration $A$V$_{3}$Sb$_{5}$ ($A$ \!=\! K, Rb, Cs) crystallize in the space group $P6/mmm$ (No. $191$) and the point group $C_{6v}$. The crystal structure has a stacking order of A-Sb1-VSb2-Sb1-A, as shown in Figs.~\ref{Fig1}(a),(b). First-principles density functional theory calculations suggest that the Fermi level of these compounds is populated dominantly by V orbitals~\cite{Hu_npjQM2023,Wilson_NatRevMater2024,Hao_PRB2022}. The V atoms form a two-dimensional kagom\'e lattice, composed of corner-sharing triangles and large hexagonal voids with three sites in the unit cell, as shown by the green atoms in Fig.~\ref{Fig1}(b). The electronic band structure of the kagom\'e lattice, shown in Fig.~\ref{Fig1}(c), features Dirac points at the $\rm K$ corners of the Brillouin zone, two Van Hove singularities with logarithmic divergence at the $\rm M$ zone edges at different energies and a dispersion-less flat band in the entire Brillouin zone, touching the top or the bottom of the dispersive bands at the center of the Brillouin zone - the $\Gamma$ point (see Fig.~\ref{Fig1}(d), and Ref.~\cite{Fu_PRL2021}). The energy dispersion relation of the dispersive bands is given by~\cite{Franz_PRB2009,Fulde_PRB2010}
\begin{equation*}
E_{k}^{1,2}  \!=\!  -t \Big( 1 \pm \sqrt{ 3+2\cos{(k_{x})}+4\cos{(k_{x}/2)}\cos{(\sqrt{3}k_{y}/2)}  } \Big),
\end{equation*}
while that of the flat band is $E_{k}^{3}\!=\!2t$, where $t$ is the tight-binding parameter, $k_{x}$ and $k_{y}$ are wave vectors along $x$ and $y$ directions, respectively. Depending on the sign of $t$, the flat band appears at the top or the bottom of the dispersive band edge. The flat band originates due to destructive quantum phase interference, resulting in the confinement of electronic wave functions within the hexagonal plaquettes. The density of electronic states, shown in Fig.~\ref{Fig1}(d), also reveals the existence of the flat band and the Van Hove singularities.

It is interesting to recall that the kagom\'e lattice at some fillings supports strongly correlated electrons~\cite{Fulde_PRB2010,Zhou_PRB2023}. In multi-orbital systems such as $A$V$_{3}$Sb$_{5}$, it is imperative to take into consideration that topological flat bands hosting such strongly correlated electrons can lead to exotic phenomena such as fractional quantum Hall effect and enhanced attractive pairing interaction~\cite{Okamoto_CommunPhys2022,Peotta_NComm2015,Bernevig_PRL2020}. In the present study, we set our global chemical potential at one of the Van Hove singularities, as observed in experiments~\cite{Liu_PRX2021}, which makes it viable to, in principle, realize such exotic effects. Nonetheless, in the following, we consider that the pairing interactions are present in the system already and investigate the consequences of the competing charge and loop-current orders in the resulting superconducting state.

%------------------------------------------------------
%------------------------------------------------------
\subsection{Model Hamiltonian}
\begin{figure}[t]
\includegraphics[width=85mm]{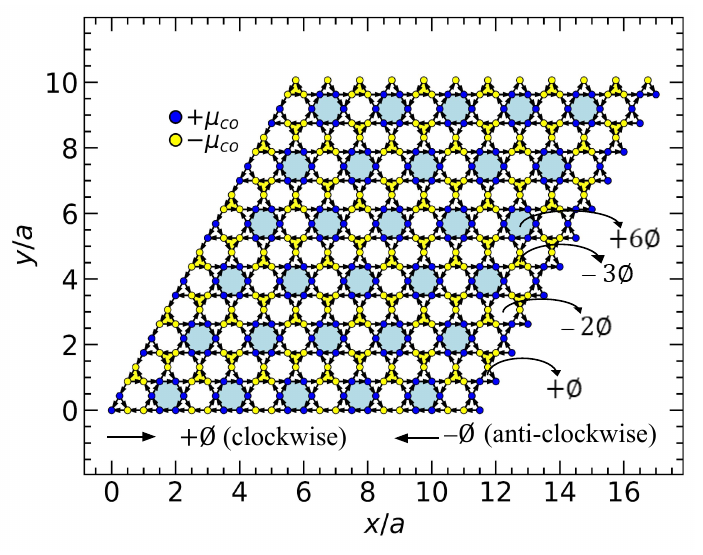}
\caption{Schematic representation of charge order in a kagom\'e lattice of size $11a\times11a$ with intertwined loop current order. The charge order has been modeled using local modulation of the chemical potential ${\mu_{co}}$, implementing the tri-hexagonal pattern as observed in scanning tunneling microscopy experiments. The arrows represent the propagation direction of the loop current and ${\phi}$ is the associated unit flux embedded through the triangular or hexagonal plaquettes due to the loop current. The charge order doubles the lattice unit cell.}
\vspace{-2mm}
\label{Fig2}
\end{figure}

\vspace{-2mm}
Two charge-order patterns have been reported---star of David and tri-hexagonal~\cite{Ortiz_PRX2021,Shumiya_PRB2021,Wang_PRB2021,Hu_PRB2022,Gupta_CommunPhys2022}; both enlarge the unit cell size to $2a\times2a$ (where $a$ is the lattice constant), confirmed by scanning tunneling microscopy and angle-resolved photoemission microscopy experiments. In particular, the tri-hexagonal charge order is dominantly observed in all three compounds and is considered to be energetically favorable. To model such a charge-ordered state, we use local modulation of the chemical potential with a $2a\times2a$ periodicity, as depicted in Fig.~\ref{Fig2}. The broken time-reversal symmetry in the charge-ordered state, in the absence of any long-range magnetic order, is incorporated into our theoretical set up by considering a loop current term in which an imaginary nearest-neighbor hopping amplitude breaks the time-reversal symmetry. It was shown in theoretical calculations that such a loop-current order, forming a chiral flux phase, is energetically favorable and it can explain the observed anomalous Hall effect in these compounds~\cite{Feng_SciBull2021}. To describe the attractive electronic interactions in the charge-ordered kagom\'e lattice, we consider the following tight-binding Hamiltonian
\begin{equation} \label{eq1}
\begin{split}
{\cal H}  = & -t\sum_{\langle i j \rangle,\sigma} (c_{i\sigma}^\dag c_{j\sigma}+{\rm H.c.} )-\sum_{i,\sigma}\left(\mu_{0}+\xi_{i}\mu_{\rm co}\right) c_{i\sigma}^\dag c_{i\sigma}\\
 & - \mathrm{i} t_{lc}\sum_{\langle i j \rangle,\sigma} (c_{i\sigma}^\dag c_{j\sigma}-{\rm H.c.}) - U_{s}\sum_{i} n_{i\uparrow}n_{i\downarrow} \\
 & - \frac{U_{t}}{2}\sum_{i,j,\sigma,\sigma^{\prime}} n_{i\sigma}n_{j\sigma^{\prime}},
\end{split}
\vspace{-2mm}
\end{equation}
where $t$ is the electron hopping energy between nearest-neighbor sites, $c_{i\sigma}^\dag$ and $c_{i\sigma}$ are the fermionic creation and annihilation operators at the site $i$ with spin $\sigma$=$\uparrow,\downarrow$, respectively, $\mu_{0}$ is the global chemical potential, $\mu_{\rm co}$ is the charge order amplitude, $\xi_{i}$ is a local variable ($\pm1$) that generates the considered tri-hexagonal charge order pattern, $\mathrm{i} t_{lc}$ is the nearest-neighbor imaginary hopping amplitude which establishes the loop-current order in the lattice and breaks the time-reversal symmetry, $n_{i\sigma} \!=\! c_{i\sigma}^\dag c_{i\sigma}$ is the electron density at the $i$$^{\rm th}$ site with spin $\sigma$, $U_s$ and $U_t$ are the strengths of the onsite and non-local attractive pair-wise interactions, respectively.

%------------------------------------------------------
%------------------------------------------------------

\subsection{Self-consistent Bogoliubov-de Gennes formalism}
\vspace{-2mm}
The attractive interaction terms in the above Hamiltonian~(\ref{eq1}) are decomposed into terms which are quadratic in fermionic operators using the standard mean-field decoupling method. The resulting Hamiltonian is given by  
\begin{align}
{\cal H}^{\rm MF}  &=  -t\sum_{\langle i j \rangle,\sigma} (c_{i\sigma}^\dag c_{j\sigma}+{\rm H.c.} )
-\sum_{i,\sigma}\left(\mu_{0}+\xi_{i}\mu_{\rm co}\right) c_{i\sigma}^\dag c_{i\sigma} \nonumber \\
& - \mathrm{i} t_{lc}\sum_{\langle i j \rangle,\sigma} (c_{i\sigma}^\dag c_{j\sigma}-{\rm H.c.})  
+ \sum_{i} (\Delta_{i}c_{i\uparrow}^\dag c_{i\downarrow}^\dag+{\rm H.c.}) \nonumber \\
&+ \frac{1}{2}\sum_{\langle ij \rangle,\sigma,\sigma^{\prime}} 
(\Delta_{ij\sigma\sigma^{\prime}}c_{i\sigma}^\dag c_{j\sigma^{\prime}}^\dag +{\rm H.c.}) 
+ \sum_{i,\sigma}\Gamma^{\rm H}_{i\sigma} c_{i\sigma}^\dag c_{i\sigma}  \nonumber \\
&+ \frac{1}{2} \sum_{\langle ij \rangle,\sigma}\Gamma^{\rm H}_{ij} c_{i\sigma}^\dag c_{i\sigma} 
-\frac{1}{2} \sum_{\langle ij \rangle,\sigma, \sigma^{\prime}}\Gamma^{\rm F}_{ij\sigma \sigma^{\prime}} c_{i\sigma}^\dag c_{j\sigma^{\prime}},
\label{H_MF}
\end{align}
where the onsite pairing amplitude $\Delta_{i}$, the off-site pairing amplitude $\Delta_{ij\sigma\sigma^{\prime}}$, the onsite Hartree potential $\Gamma^{\rm H}_{i\sigma}$, the off-site Hartree potential $\Gamma^{\rm H}_{ij}$ and the Fock potential $\Gamma^{\rm F}_{ij\sigma\sigma^{\prime}}$ are given by $\Delta_{i}\!=\!-U_s\langle c_{i\uparrow}c_{i\downarrow}\rangle$, $\Delta_{ij\sigma\sigma^{\prime}} \!=\!-(U_t/2)\langle c_{i\sigma}c_{j\sigma^{\prime}}\rangle$, $\Gamma^{\rm H}_{i\sigma} \!=\! -U_s\langle c_{i\sigma}^{\dag} c_{i\sigma}\rangle$, $\Gamma^{\rm H}_{ij} \!=\! -U_t\sum_{\sigma}\langle c_{j\sigma}^{\dag}c_{j\sigma}\rangle$ and $\Gamma^{\rm F}_{ij\sigma\sigma^{\prime}} \!=\!  -(U_t/2)\langle c_{i\sigma}^{\dag}c_{j\sigma^{\prime}}\rangle$.
The total Hamiltonian in Eq.~(\ref{H_MF}) is diagonalized by expressing it in the Bogoliubov-de Gennes (BdG) basis, using the following unitary transformation
\begin{equation} \label{eq8}
c_{i\sigma} = \sum_{n} u_{n\sigma}^{i}\gamma_{n}+v_{n\sigma}^{i\ast}\gamma_{n}^\dag ,
\end{equation}
where $\gamma_{n}^\dag$ ($\gamma_{n}$) is a fermionic creation (annihilation) operator at the $n^{\text{th}}$  eigenstate and $u_{ni}^\sigma$ ($v_{ni}^{\sigma}$) is the corresponding quasiparticle (quasihole) amplitude at site $i$ with spin $\sigma$. 

The energy eigenvalues and corresponding eigenvectors, in terms of quasiparticle and quasihole amplitudes, are then obtained by solving the BdG equations
\begin{align}
\sum_{j}{\cal H}_{ij}^{\rm BdG} \psi_n^j=E_{n} \psi_n^i,
\label{H_BdG}
\end{align}
where the BdG Hamiltonian matrix is expressed as
\begin{align}
&{\cal H}_{ij}^{\rm BdG}\!=\! \nonumber \\
&\small{
\begin{pmatrix}
\Lambda_{\uparrow\uparrow}^{ij} & \Lambda_{\uparrow\downarrow}^{ij} & \Delta_{ij\uparrow\uparrow}& \Delta_{i\uparrow\downarrow}+\Delta_{ij\uparrow\downarrow}\\[6pt]
\Lambda_{\downarrow\uparrow}^{ij} & \Lambda_{\downarrow\downarrow}^{ij} &\Delta_{i\downarrow\uparrow}+\Delta_{ij\downarrow\uparrow} & \Delta_{ij\downarrow\downarrow}\\[6pt]
\Delta_{ij\uparrow\uparrow}^{*} & \Delta_{i\downarrow\uparrow}^{*}+\Delta_{ij\downarrow\uparrow}^{*} & -\Lambda_{\uparrow\uparrow}^{ij *} & -\Lambda_{\uparrow\downarrow}^{ij *}  \\[6pt]
\Delta_{i\uparrow\downarrow}^{*}+\Delta_{ij\uparrow\downarrow}^{*} & \Delta_{ij\downarrow\downarrow}^{*} & -\Lambda_{\downarrow\uparrow}^{ij *} & - \Lambda_{\downarrow\downarrow}^{ij *} 
\end{pmatrix},
}
\end{align}
\noindent in the basis $\psi_n^i=[u_{n\uparrow}^i, u_{n\downarrow}^i, v_{n\uparrow}^i, v_{n\downarrow}^i]^T$, where $\Lambda_{\sigma\sigma^{\prime}}^{ij} = (-t-it_{lc}-\frac{1}{2}\Gamma^{\rm F}_{ij\sigma\sigma^{\prime}})(1-\delta_{ij})\delta_{\sigma\sigma^{\prime}}-(\mu_{0}+\xi_{i}\mu_{\rm co}+\Gamma^{H}_{ii}+\frac{1}{2}\Gamma^{\rm H}_{ij})\delta_{ij}\delta_{\sigma\sigma^{\prime}}-\frac{1}{2}\Gamma^{\rm F}_{ij\sigma\sigma^{\prime}}(1-\delta_{ij})$, $\delta_{ij}$ and $\delta_{\sigma\sigma^{\prime}}$ are the Kronecker's delta functions, $E_{n}$ is the energy eigenvalue of the $n^{\text{th}}$ eigenstate. The local and non-local pairing amplitudes, the Hartree and Fock potential terms are expressed in terms of the Fermi function $f(E_{n})$ and the BdG quasiparticle and quasihole amplitudes as

\begin{equation}
\label{eq10}
\begin{split}
\Delta_{i} & =-U_s\langle c_{i\uparrow}c_{i\downarrow} \rangle\\ 
 & =-U_s \sum_{n} \left[u_{n\uparrow}^{i} v_{n\downarrow}^{i\ast}(1-f(E_n))+u_{n\downarrow}^{i} v_{n\uparrow}^{i\ast}f(E_n) \right]\\
& =-\frac{U_s}{2} \sum_{n} \left[u_{n\uparrow}^{i} v_{n\downarrow}^{i\ast}- u_{n\downarrow}^{i} v_{n\uparrow}^{i\ast} \right] \tanh\left( \frac{E_n}{2k_{B}T} \right),
\end{split}
\end{equation}
\begin{equation}
\label{eq11}
\begin{split}
\Delta_{ij\sigma\sigma^{\prime}} & =-\frac{U_t}{2}\langle c_{i\sigma}c_{j\sigma^{\prime}} \rangle \\
& =-\frac{U_t}{2} \sum_{n} \left[u_{n\sigma}^{i} v_{n\sigma^{\prime}}^{j\ast}(1-f(E_n))+u_{n\sigma^{\prime}}^{j}v_{n\sigma}^{i\ast}f(E_n) \right]\\
& =-\frac{U_t}{4} \sum_{n} \left[u_{n\sigma}^{i} v_{n\sigma^{\prime}}^{j\ast}- u_{n\sigma^{\prime}}^{j} v_{n\sigma}^{i\ast} \right] \tanh\left( \frac{E_n}{2k_{B}T} \right),
\end{split}
\end{equation}
\begin{equation}
\label{eq12}
\begin{split}
\Gamma^{\rm H}_{i,\sigma} & = -U_s\langle c_{i\sigma}^{\dag} c_{i\sigma} \rangle \\
& = -U_s \sum_{n} |v_{n\sigma}^{i}|^{2} \left[(1-f(E_n))+|u_{n\sigma}^{i}|^{2} f(E_n) \right]\\
& = -\frac{U_s}{2} \sum_{n} \left[|v_{n\sigma}^{i}|^{2}- |u_{n\sigma}^{i}|^{2} \right] \tanh\left( \frac{E_n}{2k_{B}T} \right),
\end{split}
\end{equation}
\begin{equation}
\label{eq13}
\begin{split}
\Gamma^{\rm H}_{ij} & = -U_t\sum_{\sigma}\langle c_{j\sigma}^{\dag}c_{j\sigma} \rangle \\
& = - U_t \sum_{\sigma}\sum_{n} \left[|v_{n\sigma}^{j}|^{2}(1-f(E_n))+|u_{n\sigma}^{j}|^{2}f(E_n) \right]\\
& = -\frac{U_t}{2} \sum_{n,\sigma} \left[|v_{n\sigma}^{j}|^{2}- |u_{n\sigma}^{j}|^{2} \right] \tanh\left( \frac{E_n}{2k_{B}T} \right),
\end{split}
\end{equation}
\begin{equation}
\label{eq14}
\begin{split}
\Gamma^{\rm F}_{ij\sigma\sigma^{\prime}} & =  -\frac{U_t}{2}\langle c_{i\sigma}^{\dag}c_{j\sigma^{\prime}}\rangle\\
& = -\frac{U_t}{2} \sum_{n} \left[v_{n\sigma}^{i} v_{n\sigma^{\prime}}^{j\ast}(1-f(E_n))+u_{n\sigma}^{i\ast} u_{n\sigma^{\prime}}^{j}f(E_n) \right]\\
& = -\frac{U_t}{4} \sum_{n} \left[v_{n\sigma}^{i} v_{n\sigma^{\prime}}^{j\ast}-u_{n\sigma}^{i\ast} u_{n\sigma^{\prime}}^{j} \right] \tanh\left( \frac{E_n}{2k_{B}T} \right),
\end{split}
\end{equation}
where $f(E_n) = 1/(1+\exp(E_{n}/{k_{B}T}))$ is the Fermi-Dirac distribution function, which satisfies the following relations: $\langle \gamma_{n}^{\dag}\gamma_{n}\rangle = f(E_n)$ and $\langle \gamma_{n}\gamma_{n}^{\dag} \rangle = 1-f(E_n)$, $T$ is the temperature and $k_{B}$ is the Boltzmann constant. 

\subsection{Calculation of observables}
\vspace{-2mm}
In what follows, we perform an iterative procedure to compute the quantities in Eqs.~(\ref{eq10})-(\ref{eq14}) using the BdG equations (\ref{H_BdG}). The iterations are continued until self-consistency is achieved for these five quantities at each lattice site and for the nearest-neighbor bonds.\\

\noindent \textit{I. Pairing amplitudes:} The superconducting order parameters, the pairing amplitudes in the onsite and the nearest-neighbor (NN) channels, are obtained from the converged eigenvalues and eigenvectors as\\

\noindent (a) onsite singlet: $\Delta_{i}^{s,{\rm on}} = U_s\langle c_{i\uparrow}c_{i\downarrow} \rangle$,\\
\noindent (b) NN singlet: $\Delta_{ij}^{s,{\rm NN}}=  -\frac{U_t}{2}\langle c_{i\downarrow}c_{j\uparrow}- c_{i\uparrow}c_{j\downarrow}\rangle$,\\
\noindent (c) NN equal-spin triplet: $\Delta_{ij\sigma\sigma}^{t,{\rm NN}}=  -U_t\langle c_{i\sigma}c_{j\sigma}\rangle$,\\
\noindent (d) NN opposite-spin triplet:\! $\Delta_{ij\uparrow\downarrow}^{t,{\rm NN}}\!=\!-\frac{U_t}{2}\langle c_{i\downarrow}c_{j\uparrow}\!-\!c_{i\uparrow}c_{j\downarrow}\rangle$.\\

\noindent \textit{II. Pair-pair correlation function:} The nature of the superconducting state, particularly the momentum dependence, can be studied using the Fourier transformation of the pair-pair correlation function, which is given by
\begin{equation}
C({\bf q}) = \frac{1}{N}\sum_{i,j}\langle \Delta_i \Delta_j \rangle e^{-i{\bf q} \cdot {\bf r}_{ij}},
\end{equation}
where ${\bf q}$ is the momentum, $\Delta_i$ and $\Delta_j$ are onsite pairing amplitudes at the lattice sites $i$ and $j$ and ${\bf r}_{ij}$ is the distance between these sites. In the below results, we present the intensity of the above pair-pair correlation using $I({\bf q})=|C({\bf q})|^2$, to minimize undesired Fourier components away from the characteristic momenta. We obtain this observable also for the NN pairing amplitudes using a similar method as above.\\

\noindent \textit{III. Local density of states:} The local density of states, a useful observable that can be compared with the tunneling conductance spectrum, is calculated using 
\begin{equation}
\label{eq15}
\begin{split}
\rho_{i}(E) \!=\! \sum_{n} \left[|u_{n\sigma}^{i}|^{2}\delta(E-E_{n})+ |v_{n\sigma}^{i}|^{2} \delta(E+E_{n})\right],
\end{split}
\end{equation}
\noindent where $\delta(E\pm E_n)$ are the Dirac delta functions that are approximated by Gaussian functions in our numerical calculations. The total density of states is obtained by summing over all lattice sites.

%------------------------------------------------------
\begin{figure*}
\includegraphics[width=155mm]{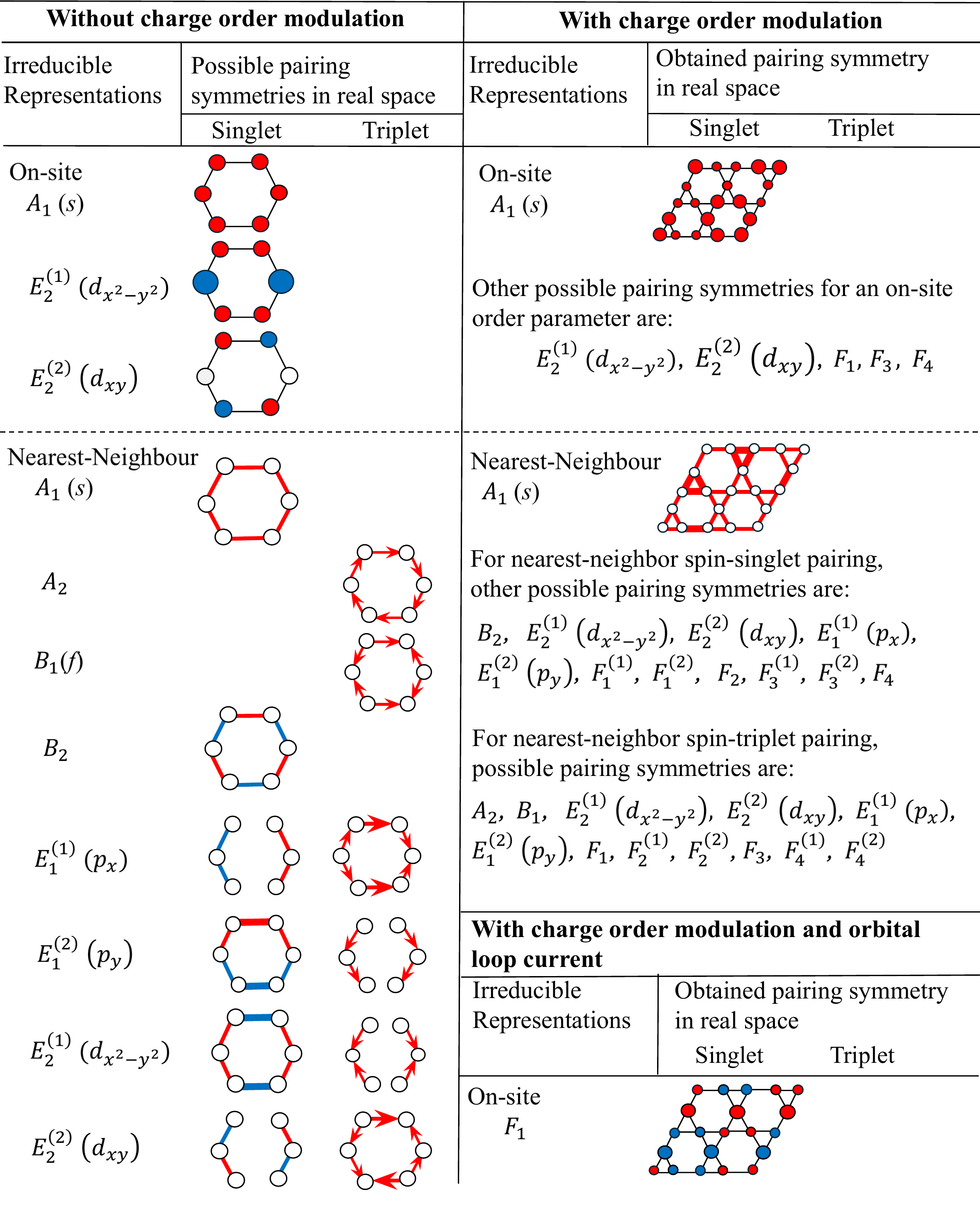}
\caption{Summary of different onsite and nearest-neighbor pairing states on the kagom\'e lattice. For singlet states, the relative sign and the strength of the pairing interactions are represented by the color  (red: positive, white: zero, blue: negative) and the size of circles or the width of the bonds. In the cases of triplet pairings, the red color and the arrow orientation indicate positive sign. Arrow thickness indicates the strength. For the nearest-neighbor pairing, $\Delta_{ij} \!=\! \Delta_{ji}$ for $s$-wave and $d+id$ pairing symmetries and $\Delta_{ij} \!=\! -\Delta_{ji}$ for $p$$\pm$$ip$ and $f$-wave pairing symmetries due to their antisymmetric nature under exchange of real-space indices.}
%\vspace{-0mm}
\label{Fig3}
\end{figure*} 
%------------------------------------------------------
%------------------------------------------------------
\begin{figure*}
\includegraphics[width=165mm]{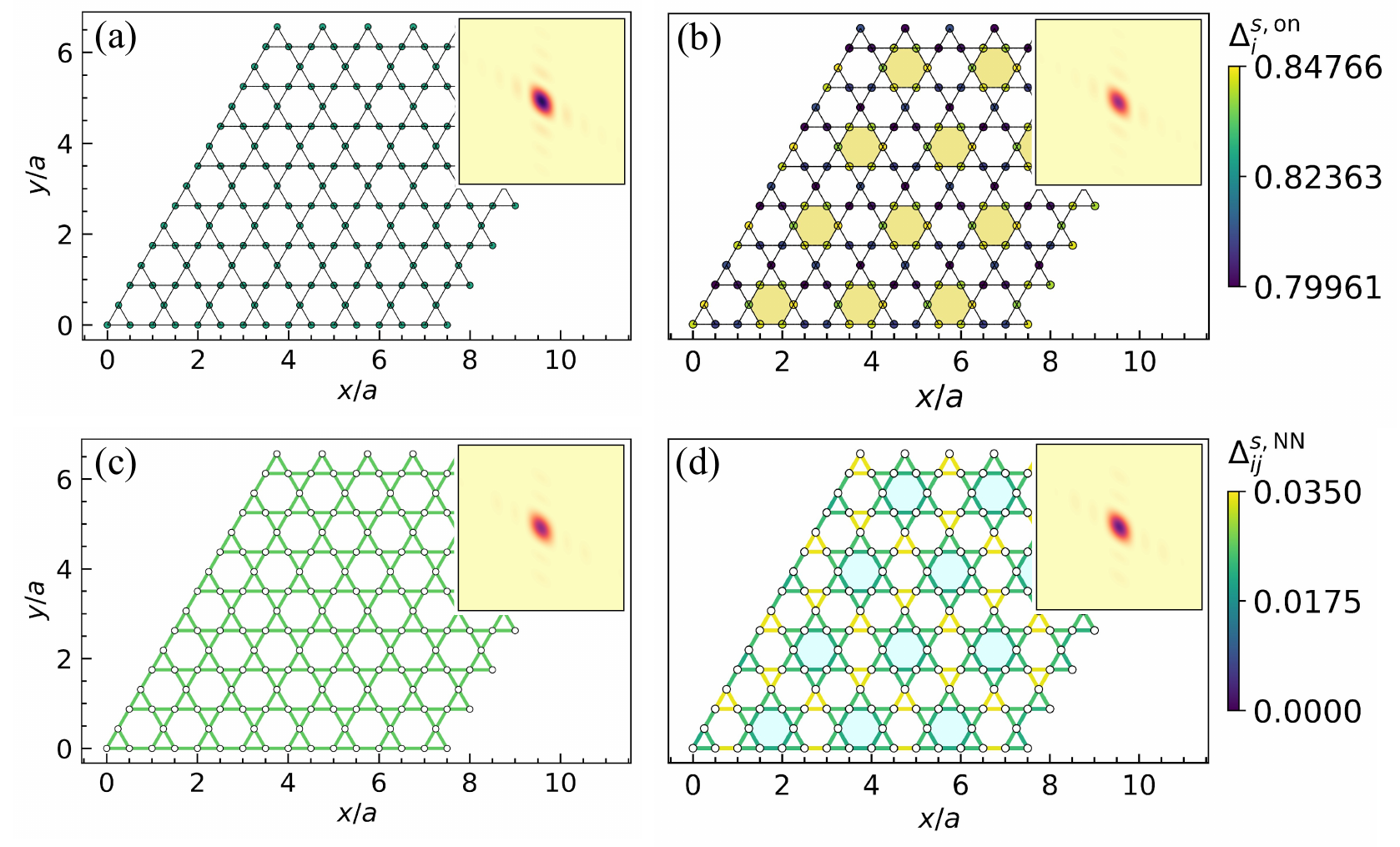}
\caption{Profiles of (a)-(b) the onsite singlet pairing amplitude $\Delta_i^{s, {\rm on}}$ and (c)-(d) the nearest-neighbor (NN) singlet pairing amplitude $\Delta_{ij}^{s, {\rm NN}}$, in the absence (left column) and presence (right column) of the charge order modulation. The order parameters were obtained self-consistently on a $7a\times 7a$ lattice with periodic boundary conditions. The onsite pairing amplitudes are shown by the colored circles and the NN pairing amplitudes by the colored bonds. (a) and (c) show uniform $s$-wave pairing symmetries both in onsite and NN channels in the absence of the charge order modulation \textit{i.e.} with $\mu_{\rm co}\!=\!0$, while (b) and (d) show pair density wave orders with $\mu_{\rm co}\!=\!t/3$. The shaded regions in (b) and (d) indicate the hexagonal blocks on which the tri-hexagonal charge order has been incorporated. Other parameters used are $U_s = 2t$ and $U_t = 2t/3$. Insets show the colormap of the intensity of the pair-pair correlation function $I({\bf q})$ in the momentum range $[q_x,q_y]=[1.3\pi/a,1.3\pi/a]$.}
%\vspace{-2mm}
\label{Fig4}
\end{figure*} 
%------------------------------------------------------
%------------------------------------------------------
\begin{figure*}
\centering
\includegraphics[width=165mm]{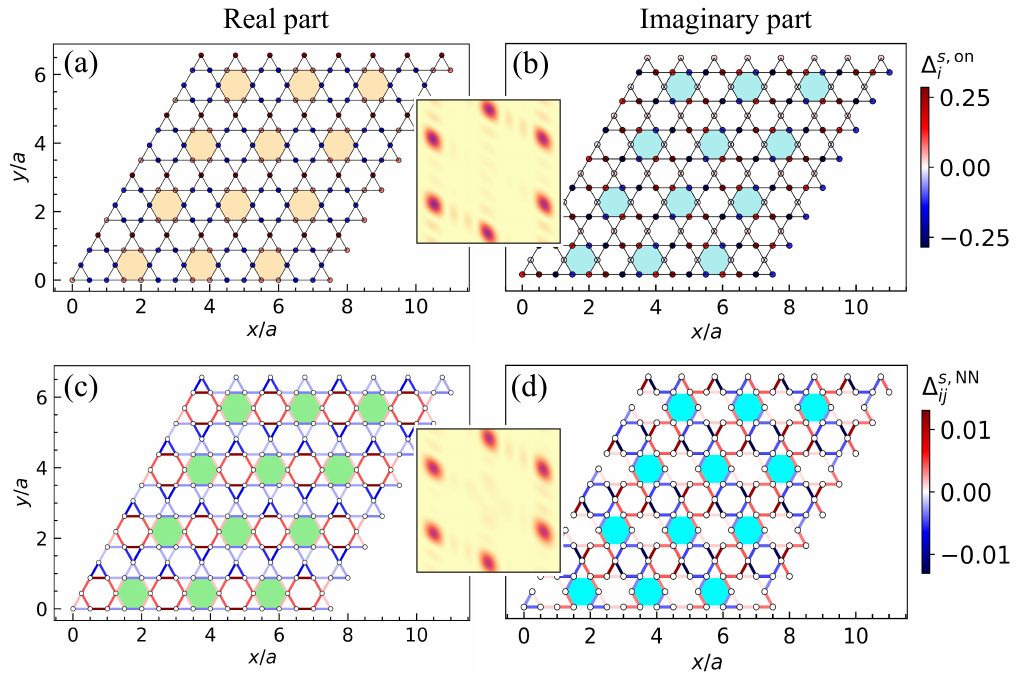}
\caption{Profiles of (a)-(c) the real part and (b)-(d) the imaginary part of the onsite singlet pairing amplitude $\Delta_{i}^{s, {\rm on}}$ (top row) and the nearest-neighbor (NN) singlet pairing amplitude $\Delta_{ij}^{s, {\rm NN}}$ (bottom row) in the presence of both charge order and loop current. The order parameters were obtained self-consistently on a $7a\times 7a$ lattice with periodic boundary conditions. The onsite pairing amplitudes are shown by the colored circles and the NN pairing amplitudes by the colored bonds. The shaded regions in (b) and (d) indicate the hexagonal blocks on which the tri-hexagonal charge order has been incorporated. Insets show the colormap of the intensity of the pair-pair correlation function $I({\bf q})$ in the momentum range $[q_x,q_y]=[1.3\pi/a,1.3\pi/a]$, computed using the complex order parameters. The six-peak structure of $I({\bf q})$ indicates the realization of a chiral pair density wave state in the presence of the loop current. Other parameters used are $U_s = 2t$, $U_t = 2t/3$, $\mu_{co} = t/3$ and $t_{lc} = t$.} 
%\vspace{-2mm}
\label{Fig5}
\end{figure*} 
%------------------------------------------------------
%\subsection{Numerical parameters}
We use, throughout this paper, lattice spacing $a \!=\! 1$, $\hbar \!=\! 1$, $\mu_{B} \!=\! 1$ and expressed the energies in units of the hopping energy $t$, which is set to $t \!=\! 1$. The global chemical potential $\mu_0$ is kept at zero, therefore, bringing the Fermi level close to one Van Hove singularity. We use charge order amplitude $\mu_{\rm co} \!=\! t/3$, onsite attractive singlet pairing potential $U_s\!=\!2t$, off-site attractive triplet pairing potential $U_t\!=\!2t/3$, without losing our general conclusions. The temperature was kept at a low value, given by $k_BT\!=\!10^{-7}t$.

\section{Symmetry allowed pairing states}\label{secIII}
The superconducting pairing symmetries on the kagom\'e lattice can be classified based on the irreps of the point group $C_{6v}$~\cite{Liu_PRB2024}. It contains twelve elements in six conjugacy classes. The character table of the $C_{6v}$ point group is presented in Table \ref{table:I}.
%\textbf{TABLE I: The character table of the $C_{6v}$ point group}.
%\vspace{-0.1mm}
%------------------------------------------------------
\begin{table}[h!]
\caption{Character table of the point group $C_{6v}$.}
\begin{center}
\begin{tabular}{
>{\raggedleft}p{1cm} 
>{\raggedleft}p{1cm} 
>{\raggedleft}p{1cm} 
>{\raggedleft}p{1cm} 
>{\raggedleft}p{1cm} 
>{\raggedleft}p{1cm} 
>{\raggedleft\arraybackslash}p{1cm}}
 \hline\hline
 $C_{6v}$  &  
 I & $2C_{6}^{\pm}$  & $2C_{3}^{\pm}$  &  $C_{2}$  & $3\sigma_{v}$  & $3\sigma_{d}$\\ [0.5ex] 
 \hline
 $A_1$  &  1  &  1  &  1 &  1  &  1  &  1\\ 
 $A_2$ & 1 & 1 & 1 & 1 & -1 & -1\\
 $B_1$ & 1 & -1 & 1 & -1 & 1 & -1\\
 $B_2$ & 1 & -1 & 1 & -1 & -1 & 1\\
 $E_1$ & 2 & 1 & -1 & -2 & 0 & 0\\
 $E_2$ & 2 & -1 & -1 & 2 & 0 & 0\\[1ex] 
 \hline\hline
\end{tabular}
\end{center}
\label{table:I}
\end{table}
%------------------------------------------------------
%------------------------------------------------------
\begin{table}[h!]
\caption{Character table of the point group $C_{6v}^{'''}$.}
\begin{center}
\begin{tabular}{
>{\raggedleft}p{0.6cm} 
>{\raggedleft}p{0.6cm} 
>{\raggedleft}p{0.6cm} 
>{\raggedleft}p{0.6cm} 
>{\raggedleft}p{0.8cm} 
>{\raggedleft}p{0.6cm} 
>{\raggedleft}p{0.6cm} 
>{\raggedleft}p{0.6cm} 
>{\raggedleft}p{0.7cm} 
>{\raggedleft}p{0.6cm}
>{\raggedleft\arraybackslash}p{0.6cm}}
 \hline\hline
 $C_{6v}$  &  I  & $t_{i}$ & $C_{2}$  & $t_{i}C_{2}$ & $C_{3}$  &  $C_{6}$  & $\sigma_{v}$  & $t_{i}\sigma_{v}$ & $\sigma_{d}$ & $t_{i}\sigma_{d}$\\ [0.5ex] 
 \hline
 $A_1$  &  1  &  1  &  1 &  1  &  1  &  1 & 1 & 1 & 1 & 1\\ 
 $A_2$ & 1 & 1 & 1 & 1 & 1 & 1 & -1 & -1 & -1 & -1\\
 $B_1$ & 1 & 1 & -1 & -1 & 1 & -1 & 1 & 1 & -1 & -1\\
 $B_2$ & 1 & 1 & -1 & -1 & 1 & -1 & -1 & -1 & 1 & 1\\
 $E_1$ & 2 & 2 & -2 & -2 & -1 & 1 & 0 & 0 & 0 & 0\\
 $E_2$ & 2 & 2 & 2 & 2 & -1 & -1 & 0 & 0 & 0 & 0\\
 
 $F_1$ & 3 & -1 & 3 & -1 & 0 & 0 & 1 & -1 & 1 & -1\\
 
 $F_2$ & 3 & -1 & 3 & -1 & 0 & 0 & -1 & 1 & -1 & 1
 \\
 $F_3$ & 3 & -1 & -3 & 1 & 0 & 0 & 1 & -1 & -1 & 1\\
 
 $F_4$ & 3 & -1 & -3 & 1 & 0 & 0 & -1 & 1 & 1 & -1\\[1ex] 
 \hline\hline
\end{tabular}
\end{center}
\label{table:II}
\vspace{-4mm}
\end{table}
%------------------------------------------------------
The set denoted by I contains the identity, $2C_{6}^{\pm}$ denotes a rotation by $\pm \pi/3$ around the principal axis (the axis perpendicular to the plane of the two-dimensional lattice), $2C_{3}^{\pm}$ contains a rotation by $\pm 2\pi/3$ around the principal axis and $C_{2}$ contains a rotation by $\pi$ around the principal axis. It also includes two inequivalent mirror planes, $\sigma_{v}$ and $\sigma_{d}$. There are six inequivalent irreps for the $C_{6v}$ point group: four of the irreps ($A_1$, $A_2$, $B_1$ and $B_2$) are one-dimensional and two of the irreps ($E_1$ and $E_2$) are two-dimensional. In the presence of the charge order and loop current order, the unit cell of the underlying lattice doubles. This system with broken time-reversal symmetry and a $2a\times2a$ unit cell, is described by the extended point group $C_{6v}^{'''}$~\cite{Wagner_PRB2023,Holbaek_PRR2025}. It contains four one-dimensional irreps ($A_1$, $A_2$, $B_1$, $B_2$), two two-dimensional irreps ($E_1$, $E_2$) and four additional three-dimensional irreps $F_{i}$, $i = 1,...,4$. Although, the one- and two-dimensional irreps of the point group $C_{6v}$ preserve the translational symmetry of the original kagom\'e lattice, the irreps of the extended point group $C_{6v}^{'''}$ lead to a $2a\times2a$ enhancement in the unit cell, which explicitly breaks the translational symmetry. These irreps of the extended point group $C_{6v}^{'''}$ are shown in the character table~\ref{table:II}. The real-space structures of these states are schematically presented in Fig.~\ref{Fig3} for three different cases: without any charge order, with only charge order modulation and with both charge order and the loop current order. For the latter case, the real-space representations of the pairing symmetries are elaborated in the Appendix~\ref{App_A}.

\section{Real-space pairing profiles}\label{secIV}
Having discussed the various types of possible pairing symmetries, which could occur on the kagom\'e lattice in the absence or presence of the charge order and the loop current, we now investigate the real-space profiles of the pairing amplitudes in different pairing channels, obtained via our self-consistent BdG calculations. In this study, we use relatively smaller lattice sizes to enable a clear visualization of the real-space configurations of the pairing amplitudes. We use periodic boundary conditions to minimize finite-size effects, that allow us to model the behavior of a large crystal with a relatively
small lattice.

\subsection{Pairing in the absence of the loop current}
We begin our efforts by considering the simplest situation in our model Hamiltonian, by setting the charge order and loop current terms to zero, \textit{i.e.} $\mu_{\rm co}\!=\!0$ and $t_{\rm lc}\!=\!0$. In this case, a uniform $s$-wave pairing symmetry, which belongs to $A_1$ irreducible representation (irrep), is realized in both onsite and nearest-neighbor channels, as shown in Figs.~\ref{Fig4}(a) and (c). In the presence of only the charge order, \textit{i.e.} with $\mu_{\rm co}\! \neq \!0$, but with $t_{\rm lc}\!=\!0$, the singlet pairing amplitudes in both the onsite and NN channels reveal periodic modulations, as shown in Figs.~\ref{Fig4}(b) and (d). In this case, there is a unit-cell doubling due to the charge order and the $A_1$-symmetric hexagonal block in the larger unit cell repeats with a $2a\times2a$ periodicity. In all four cases, however, the pairing occurs at zero characteristic momentum, as revealed by the intensity of the Fourier transform of the pair-pair correlation function $I({\bf q})$, depicted in the insets of Figs.~\ref{Fig4}(a)-(d).

\subsection{Pairing in the presence of the loop current}
The periodic flux modulation in the presence of the loop current (shown in Fig.~\ref{Fig2}) dramatically modifies the nature of the superconducting pairing in the kagom\'e lattice. As shown in our previous study~\cite{Mohanta_PRB2023}, the resulting superconducting state is a chiral pair density wave of Cooper pairs in the onsite pairing channel, having a complex pairing amplitude. Here we find that a similar pair density wave state is realized in both onsite and the nearest-neighbor channels, as shown by the real and imaginary parts of the pairing amplitudes in Fig.~\ref{Fig5}(a)-(d). The insets, showing the intensity of the Fourier transform of the pair-pair correlation function $I({\bf q})$ which is computed using the complex pairing amplitudes, reveal a six-peak structure with three characteristic momenta in the resulting superconducting state. The pairing profiles reveal $d_{x^2-y^2}\!+\!id_{xy}$-wave symmetry of $2a\times2a$ periodicity in both onsite and nearest-neighbor pairing channels. The real part of the onsite singlet pairing symmetry in the doubled unit cell matches with the symmetry of the $F_{1}$ irrep which breaks the lattice translational symmetry, as discussed in Appendix~\ref{App_A}.

%------------------------------------------------------
\begin{figure}
\includegraphics[width=85mm]{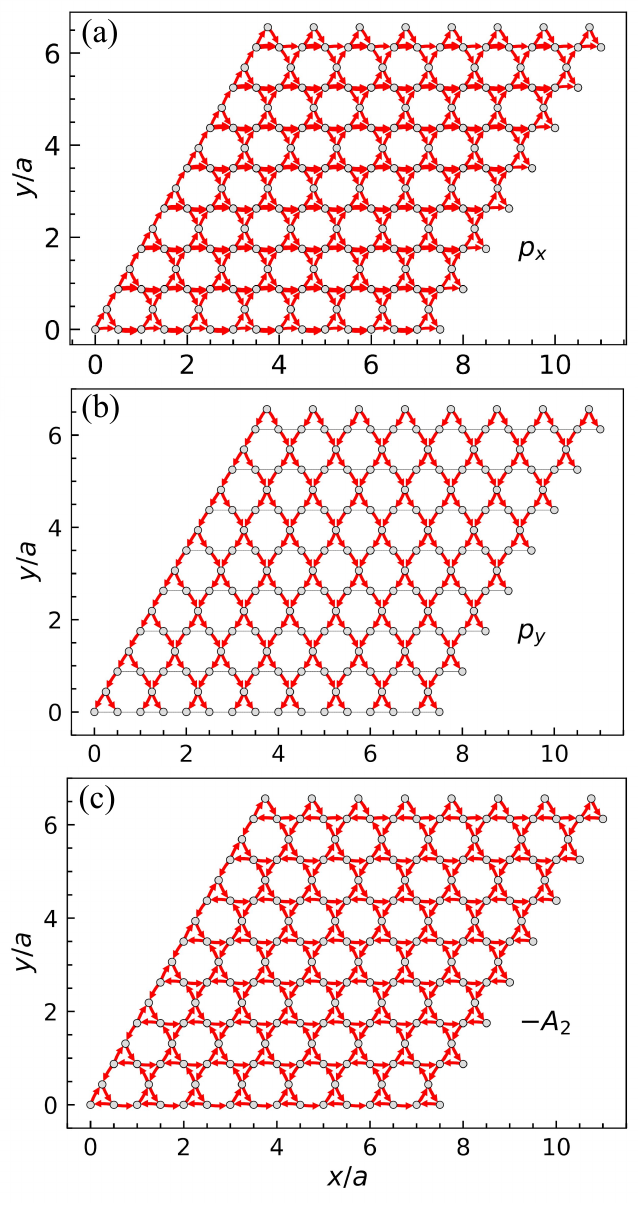}
\caption{Schematic representations of (a) $p_x$ wave, (b) $p_y$ wave and (c) $-A_2$ symmetry state in the nearest-neighbor triplet channel---the pairing symmetries that are realized in the presence of the Rashba spin-orbit coupling and absence of the charge order and loop current. The directions of the arrows represent the signs of the nearest-neighbor triplet pairing amplitudes, as schematized in Fig.3. In (a), the thicker arrows along the $+x$ direction represent pairing amplitude of double value.} 
\vspace{-2mm}
\label{Fig6}
\end{figure} 
%------------------------------------------------------

\subsection{Pairing in the presence of Rashba spin-orbit coupling}
In the thin film geometry, the broken structural inversion symmetry at the interface between $A$V$_3$Sb$_5$ and the substrate, Rashba spin-orbit coupling can arise. To investigate the effect of the Rashba spin-orbit coupling on the superconducting pairing symmetry, we include in our Hamiltonian in Eq.~(\ref{eq1}) an additional term
\begin{equation}
\label{eq16}
\begin{split}
H_{\rm RSOC} = -\frac{\mathrm{i}\lambda}{2a}\sum_{\langle ij \rangle,\sigma,\sigma\prime} (\boldsymbol{\sigma} \times \hat{d}_{ij})^{z}_{\sigma\sigma\prime} c_{i\sigma}^{\dag} c_{j\sigma^\prime},
\end{split}
\end{equation}
where $\lambda$ is the strength of the Rashba spin-orbit coupling, $\mathrm{i}$ represents the imaginary number, $\hat{d}_{ij}$ is a nearest-neighbor unit vector from site $i$ to site $j$ and $\boldsymbol{\sigma}$ represents the Pauli matrices. We carry out self-consistent BdG calculations as above to obtain the pairing amplitudes both in singlet and triplet channels, in the absence and presence of the charge order and loop current. The above Rashba spin-orbit coupling Hamiltonian represents an effective imaginary hopping in the NN channel with a spin flip. Hence, the Rashba spin-orbit coupling is found to have a strong interplay with the loop current term which also has an imaginary hopping in the NN channel but without any spin flip. In the absence of the charge order and loop current, the induced triplet pairing states are shown schematically in Fig.~\ref{Fig6}. The symmetry of the induced triplet pairing is $p_x\pm ip_y$ wave in the equal-spin channels, which belongs to the $E_1$ irrep (`+' sign is for the $\uparrow \uparrow$ pairing channel and `-' sign is for the $\downarrow \downarrow$ pairing channel) and of symmetry $-A_2$ irrep in the opposite-spin $\uparrow \downarrow$ channel. In the presence of the charge order and loop current, singlet $d_{x^2-y^2}+id_{xy}$-wave and triplet $p_x\pm ip_y$-wave pairing states appear in the NN channel, both of $2a\times 2a$ lattice periodicity. Depending on the relative strength $t_{lc}$ of the loop current term and that $\lambda$ of the Rashba spin-orbit coupling term, the singlet or triplet pairing state dominates. We also find that the degeneracy of the $\uparrow \uparrow$ and the $\downarrow \downarrow$ triplet pairing channels is lifted when an external magnetic field is applied perpendicular to the plane of the lattice, leading eventually to the $p_x + ip_y$ wave pairing state beyond a critical field strength.

%------------------------------------------------------
\begin{figure}[t!]
\includegraphics[width=86mm]{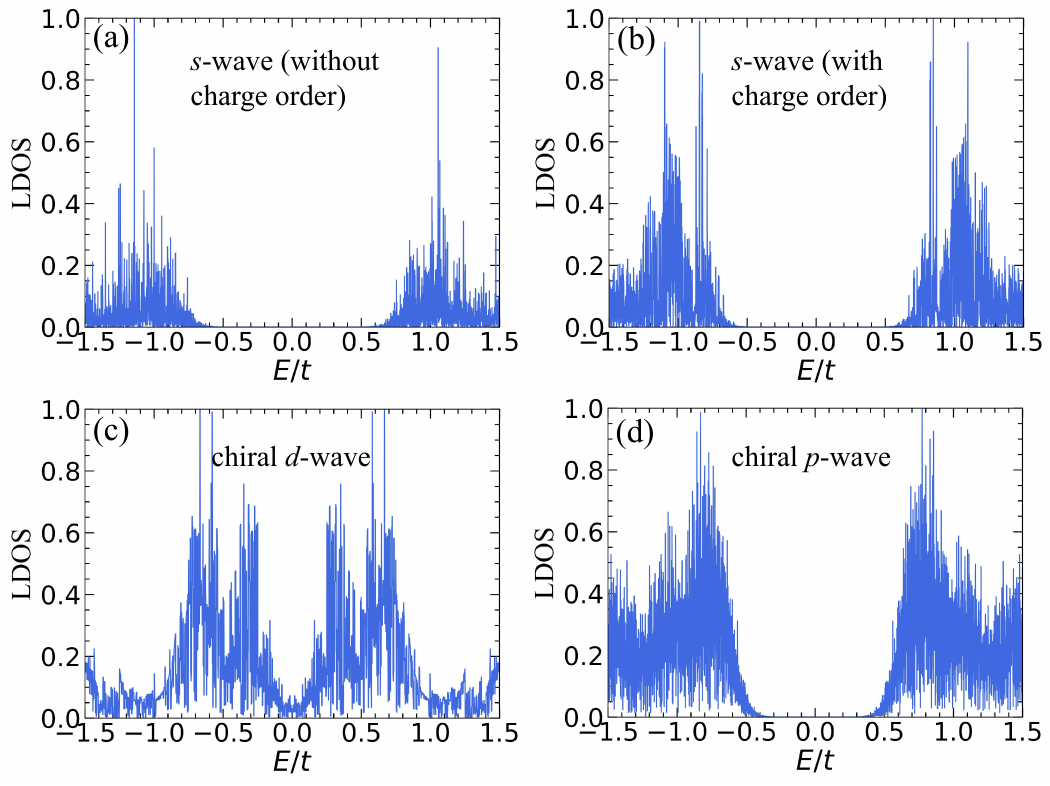}
\caption{Local density of states at a site on the kagom\'e lattice in the (a) uniform $s$-wave state realized in the absence of charge order and loop current (with $\mu_{co} = 0$ and $t_{lc} = 0$), (b) periodically-modulated $s$-wave state realized in the presence of charge order but absence of loop current (with $\mu_{co} = t/3$ and $t_{lc} = 0$), (c) chiral $d$-wave state realized in the presence of both charge order and loop current (with $\mu_{co} = t/3$ and $t_{lc} = t$), (d) chiral $p$-wave state realized in the presence of additional Rashba spin-orbit coupling and a perpendicular (with $\mu_{co} = t/3$, $t_{lc} = t$, $\lambda=0.6t$ and $B_z=0.2t$). The calculations were obtained using a $15a\times15a$ lattice with periodic boundary conditions. A clear spectral gap is present in plots (a), (b) and (d), while in-gap states forming a V-shaped density of states around zero energy is found in plot (c) for the case of a chiral $d$-wave pairing state.}
\vspace{-2mm}
\label{Fig7}
\end{figure} 
%------------------------------------------------------
\section{Local density of states for the obtained pairing symmetries}\label{secV}
The nature of the spectral energy gap in the superconducting state, which can be obtained in scanning tunneling microscopy experiments, is a useful probe for the superconducting pairing symmetry. We, therefore, calculate the local density of states, which is an equivalent theoretical observable for the tunneling conductance spectrum, in the four different pairing states realized in our model system. Figures~\ref{Fig7}(a)-(d) show the (a) uniform $s$-wave state which appears in the absence of both charge order and loop current, (b) $s$-wave state with a periodic modulation in the pairing amplitude which appears in the presence of charge order but absence of loop current, (c) $d_{x^2-y^2}+id_{xy}$-wave state which appears in the presence of both charge order and loop current, (d) $p_x+ip_y$-wave state which appears in the presence of Rashba spin-orbit coupling and a perpendicular magnetic field. A full gap is visible in the case of $s$-wave and chiral $p$-wave states (plots~\ref{Fig7}(a),(b),(d)); however, in-gap states appear in the case of $d_{x^2-y^2}+id_{xy}$-wave state (plot~\ref{Fig7}(c)) forming a V shape near zero energy. We obtain the same qualitative patterns at all three inequivalent sites of the unit cell. The finite weight of the local density of states at zero energy in the $d_{x^2-y^2}+id_{xy}$-wave state can potentially be the mechanism for the pseudogap-like feature observed in the tunneling conductance microscopy. On the other hand, the full gap in the chiral $p$-wave state can be a reminiscence of the topological superconducting state.

\section{Conclusion and outlook}\label{secVI}
Using symmetry considerations and self-consistent BdG analyses, we have established that the superconducting pairing symmetry in the charge-ordered kagom\'e superconductors $A$V$_{3}$Sb$_{5}$ with a time-reversal symmetry-breaking loop-current order is $d_{x^2-y^2}+id_{xy}$ wave in both onsite and the NN pairing channels, both of lattice periodicity $2a\times 2a$. These pairing states in onsite and NN channels constitute a chiral pair density wave which causes in-gap quasiparticle states appearing inside the bulk superconducting energy gap, resulting in a V-shaped density of states around zero energy below the superconducting transition temperature. These complex features of the density of states, which pose serious challenges to understand the nature of the superconductivity in these compounds based on scanning tunneling microscopy experiments, can be deciphered using the consideration of the formation of the chiral pair density wave state having pairing symmetries as mentioned above.

It is interesting to note that the chiral $d$-wave pairing state, which appears in the presence of the charge order and the loop current and the chiral $p$-wave pairing state, which appears in the presence of Rashba spin-orbit coupling and a Zeeman magnetic field perpendicular to the kagom\'e plane, are topological superconducting phases. Such a topological superconductor hosts at their boundaries and vortices exotic quasiparticle bound states which are considered to be useful in decoherence-free topological quantum computing and Josephson junction devices~\cite{Kheirkhah_PRB2022,Mohanta_CommPhys2023,Wang_Sciadv2023}. Hence these compounds $A$V$_3$Sb$_5$ provide a natural testbed for realizing advanced technological devices, as well as for studying unconventional emergent physical properties arising due to the interplay of various broken symmetries. Our theoretical finding of the pairing symmetries of the superconducting state serves as a stepping stone towards understanding the highly-complex superconducting state in these kagom\'e compounds.

\section*{Acknowledgements}
P.D. was supported by Ministry of Education, Government of India via a research fellowship. P.D. thanks Pankaj Sharma for discussion on BdG simulation. This work was supported by Science and Engineering Research Board, India (SRG/2023/001188) and an initiation grant (IITR/SRIC/2116/FIG). We acknowledge the National Supercomputing Mission for providing computing resources of PARAM Ganga at the Indian Institute of Technology Roorkee, which is implemented by C-DAC and supported by the Ministry of Electronics and Information Technology and Department of Science and Technology, Government of India. 

%%%%%%%%%%%%%%%%%%%%%%%%%%%%%%%%%%% Appendix material
\appendix
\refstepcounter{section}
\setcounter{figure}{0}
\renewcommand{\thefigure}{A\arabic{figure}}
\renewcommand{\theHfigure}{A\arabic{figure}}
%##################################################################
%------------------------------------------------------
\begin{figure*}
\includegraphics[width=155mm]{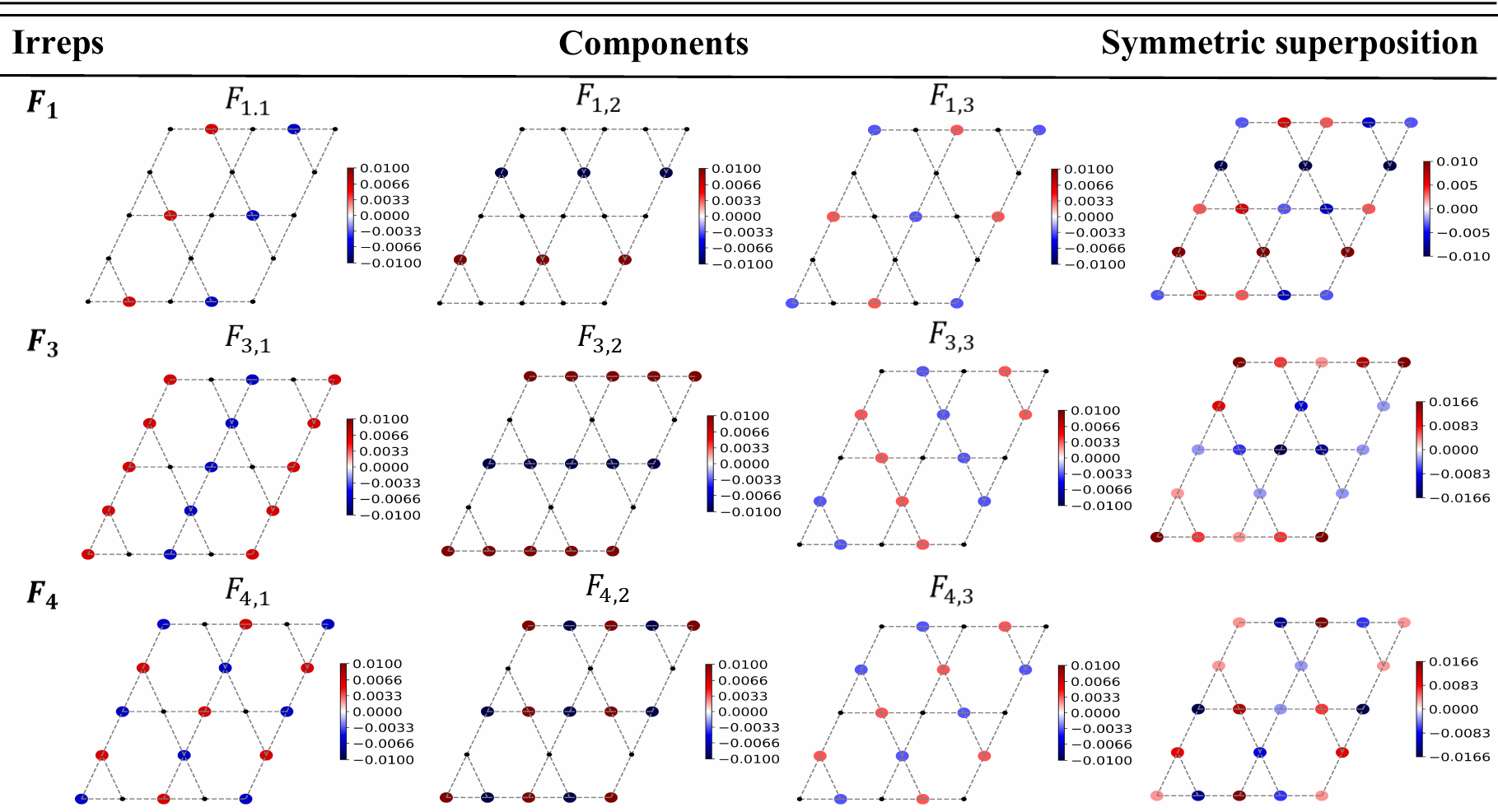}
\caption{Real-space depictions of onsite pairing states on the charge-ordered kagom\'e lattice that exhibit broken translational symmetry. The color scale represents the value of the onsite singlet pairing interaction. The first column lists the three-dimensional irreps, denoted according to the extended point group $C_{6v}^{'''}$. The second column displays each component of the irrep, while the third column illustrates a symmetric superposition of the three components.}
\label{FigA1}
\end{figure*} 
%------------------------------------------------------
%------------------------------------------------------
\begin{figure*}
\centering
\graphicspath{ {./Images/} }
\includegraphics[width=155mm]{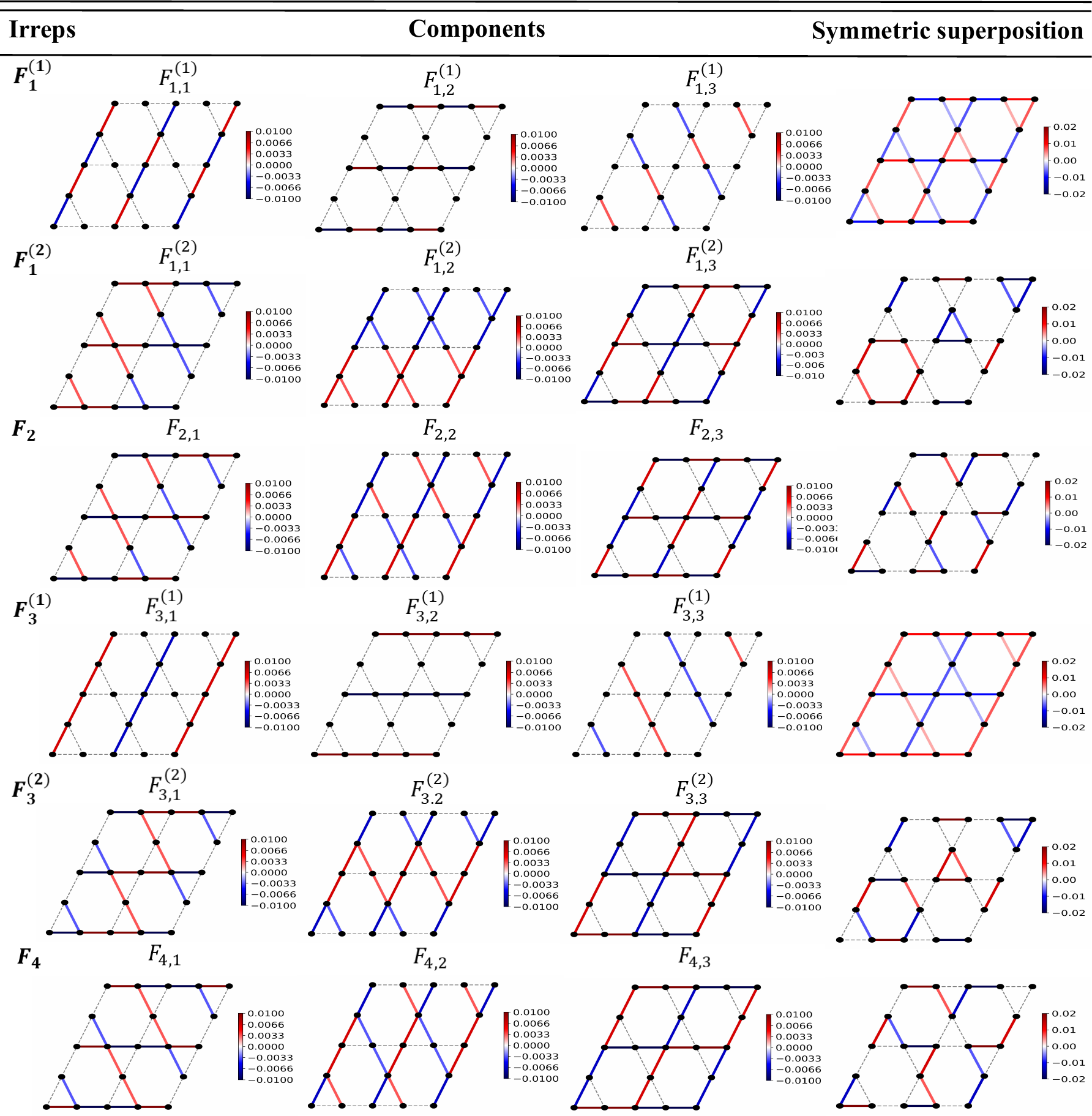}
\caption{Real-space depictions of nearest-neighbor singlet pairing states on the charge-ordered kagom\'e lattice that exhibit broken translational symmetry. The color scale represents the value of the nearest-neighbor singlet pairing interaction. The first column lists the three-dimensional
irreps, denoted according to the extended point group $C_{6v}^{'''}$ . The second column displays each component of the irreps, while the third column illustrates a symmetric superposition of the three components.}
\label{FigA2}
\end{figure*} 
%------------------------------------------------------
%------------------------------------------------------
\begin{figure*}
\centering
\graphicspath{ {./} }
\includegraphics[width=155mm]{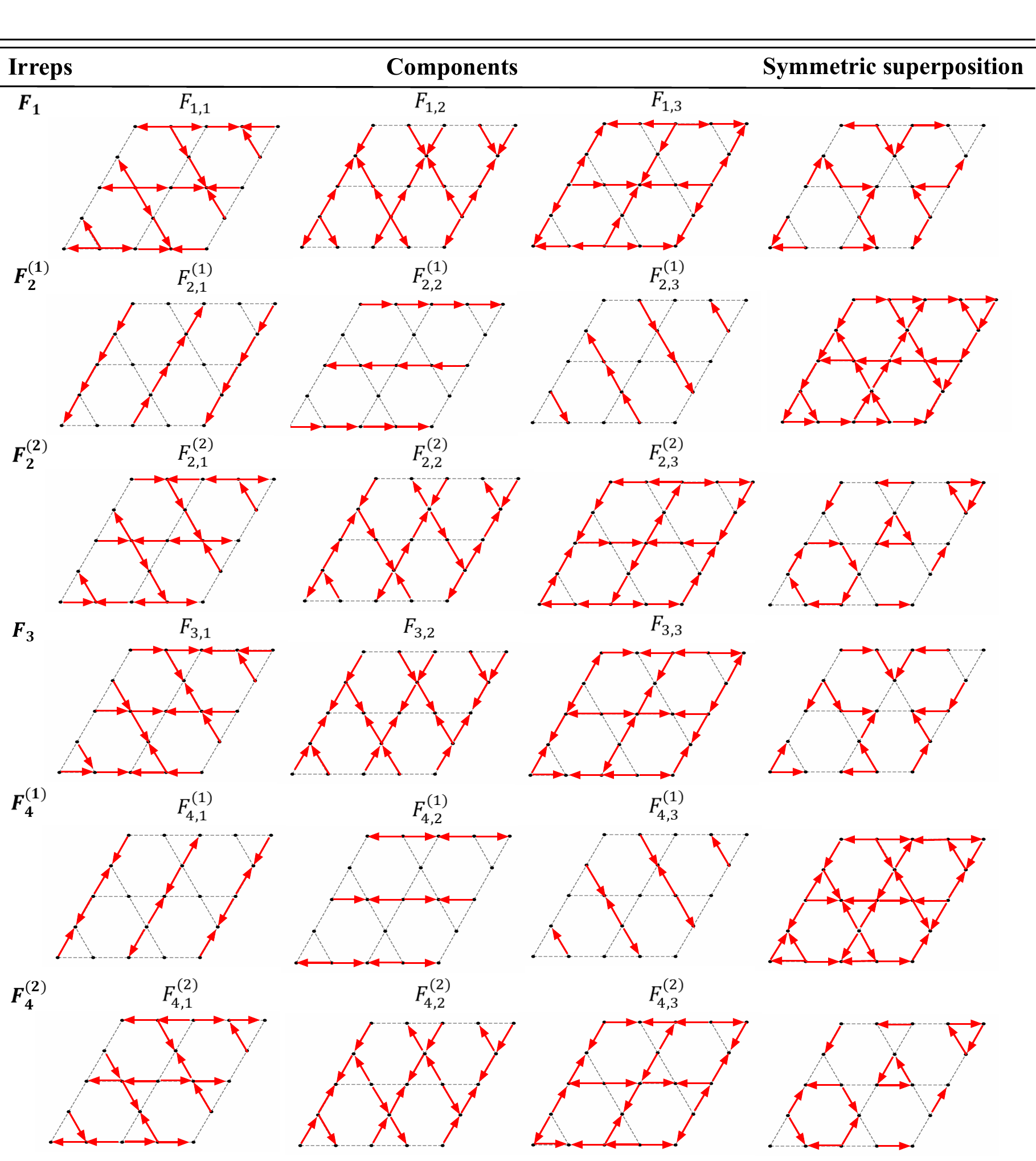}
\caption{Real-space depictions of nearest-neighbor triplet pairing states on the charge-ordered kagom\'e lattice that exhibit broken translational symmetry. The first column lists the three-dimensional
irreps, denoted according to the extended point group $C_{6v}^{'''}$ . The second column displays each component of the irreps, while the third column illustrates a symmetric superposition of the three components.}
\label{FigA3}
\end{figure*} 
%------------------------------------------------------

\vspace{2mm}
\appendix
\section{Pairing symmetries in $C_{6v}^{'''}$ point group} \label{App_A}

In this Appendix, we provide the schematic real-space configurations of various superconducting order parameters in the extended $C_{6v}^{'''}$ point group which is valid for the charge-ordered kagomé lattice with loop current~\cite{Holbaek_PRR2025}. The schematic real-space configurations of various superconducting order parameters in the original $C_{6v}$ point group, which exists in the kagom\'e lattice in the absence of the charge order and loop current, are shown in Fig.~\ref{Fig3} (left column). In the point group $C_{6v}$, the order parameters belonging to one-dimensional and two-dimensional irreps preserve translational symmetry, and are modulated within the unit cell of the original kagom\'e lattice. In the point group $C_{6v}^{'''}$, the translational symmetry is broken, resulting in a period doubling of the unit cell for the order parameters belonging to one-dimensional and two-dimensional irreps, and addition of new three-dimensional irreps $F_{i}$. The real-space profiles of these $F_{i}$ irreps, along with their decomposition of the symmetrized products, are shown in Figs.~\ref{FigA1},~\ref{FigA2} and ~\ref{FigA3}. We identified the $F_{1}$ irrep, shown in Fig.~\ref{FigA1}, in the real part of the onsite singlet pairing amplitude $\Delta_{i}^{s, {\rm on}}$, as described in Fig.~\ref{Fig5}.

%\bibliography{Ref}


\end{document}